\documentclass[aps,onecolumn,subeqns.floatfix,amsmath,showpacs]{revtex4}
\usepackage{subfigure}
\usepackage{graphicx} 
\begin{document}
 \title{Chimera states in a population of identical oscillators under planar cross-coupling}
 \author{C. R. Hens$^1$, A. Mishra $^2$,  P.K.Roy $^3$, A. Sen $^4$, S. K. Dana$^1$}
 \affiliation{$^1$CSIR-Indian Institute of Chemical Biology, Kolkata 700032, India}
 \affiliation{$^2$Department of Physics, Jadavpur University, Kolkata 700032, India}
\affiliation{$^3$Department of Mathematics, Presidency University, Kolkata 700073, India}
 \affiliation{$^4$Institute for Plasma Research, Gandhinagar 382428, India}

 \date{\today}

\begin{abstract}
We report the existence of a {\it chimera} state in an assembly of identical nonlinear oscillators that are globally linked to each
 other in a simple planar cross-coupled form. The  rotational symmetry breaking of the
 coupling term appears to be responsible for the emergence of these collective states that display a characteristic coexistence of coherent and incoherent behaviour. Our finding, seen in both a collection of van der Pol oscillators and chaotic R\"ossler oscillators, further simplifies the existence criterion for {\it chimeras} and thereby broadens the range of their applicability to real world  situations. 
\end{abstract}
\maketitle
\par Chimera states, a curious emergent behaviour of a network of coupled oscillators, has attracted a great deal of attention in recent years \cite{Kuramoto, Abrams-Strogatz,Martenes,Sen-2008,Sheeba-2010,Omelchenko-chaos,Omelchenko-2013,Sen-2013,Anna,Pikovsky}. A surprising and nonintuitive aspect of this collective state is its composition of coexisting synchronous and asynchronous behavior - an asymmetric pattern arising from a purely symmetric situation where all oscillators are identical and are  coupled symmetrically.  This spontaneous splitting of the oscillators into two subpopulations of coherent and incoherent oscillators was first discovered by Kuramoto and Battogtokh \cite{Kuramoto} for a system of phase oscillators that were coupled in a non-local fashion and subsequently studied by a number of authors including Abrams and Strogatz \cite{Abrams-Strogatz}  who named it as a {\it chimera} state. The basic interest in this state has kept growing over the years as a rich variety of such states have been discovered in various model systems \cite{Abrams-Strogatz,Martenes,Omelchenko-chaos,Omelchenko-2013,Martenes-2013} and more importantly as experimental demonstration of chimeras have also been achieved in laboratory systems \cite{Tinsley,Murphy,Maistrenko,Martenes-2013}.
Further interest has been sparked by the possibility of invoking these collective states to model such phenomena as unispheric sleep in certain mammals where one half of the brain sleeps (showing high amplitude low frequency coherent neuronal signals) while the other half is awake and displays incoherent electrical activity \cite{Levy}.
\par A key question that has received some attention in recent times is that of the basic conditions necessary for a chimera state to exist in a system of coupled oscillators. It was long assumed (on the basis of the original findings \cite{Kuramoto,Abrams-Strogatz}) that chimera states can only exist in coupled phase oscillators and only under the restrictive condition of a `non-local' coupling. More recent work has shown that
these conditions are not absolutely necessary and chimeras can exist in systems where both phase and amplitude variations are important and also in globally coupled systems \cite{Sen-2014,Schmidt}. 
 Kaneko \cite{Kaneko} observed coexisting ordered (periodic state) and disordered  (chaotic state) populations in an assembly of globally coupled chaotic maps. 
In \cite{Schmidt}, Schmidt {\it et al} demonstrated that an ensemble of Stuart-Landau oscillators with a nonlinear
global coupling could give rise to chimera states. Sethia and Sen \cite{Sen-2014} further opened up the field by showing that it was not necessary for the coupling to be nonlinear and a linear complex planar coupling was capable of producing chimera states. These chimera states had variations in both amplitudes and phases.  In this work we have tried to probe the question of the existence criterion even further 
by simplifying both the nature of the global coupling and by expanding the search for these states to ensembles of van der Pol (VDP) oscillators \cite{VDP} as well as chaotic Rossler oscillators. 

\begin{figure*} [t!]
{\includegraphics[height=7cm,width=9cm]{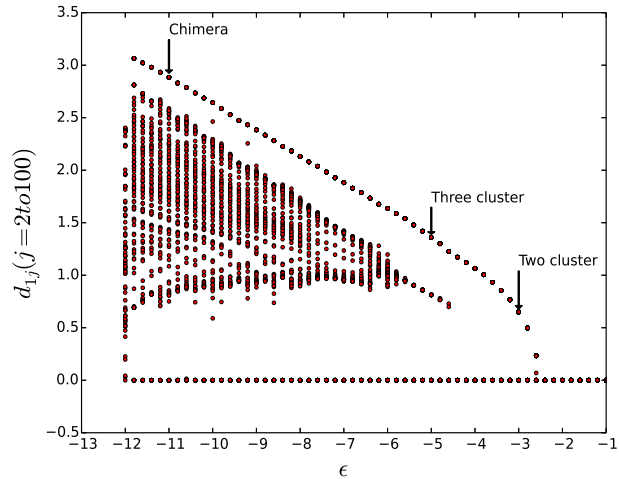}}
\caption{Network of VDP oscillators. Distance $d_{1,j}$ of oscillator-1 from all others with $\epsilon (K=0.04)$. A single cluster state bifurcate into 2-cluster ($\epsilon=2.$, then to 3-cluster and finally to chimera states. Bifurcation points= -2.6 (2-cluster), -4.6 (3-cluster).} 
\label{distance_vdp} 
\end{figure*}

\begin{figure*}  [h!]
\subfigure(a)\includegraphics[height=8cm,width=6cm]{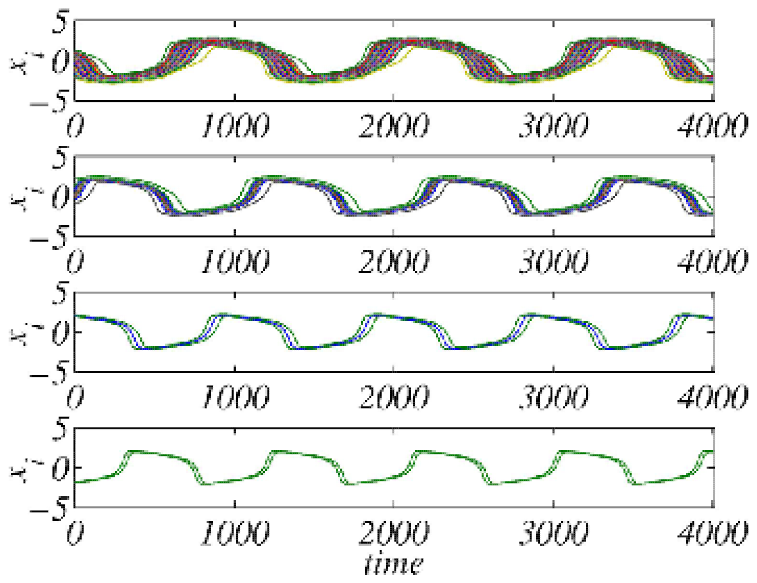}
\subfigure(b)\includegraphics[height=8cm,width=6cm]{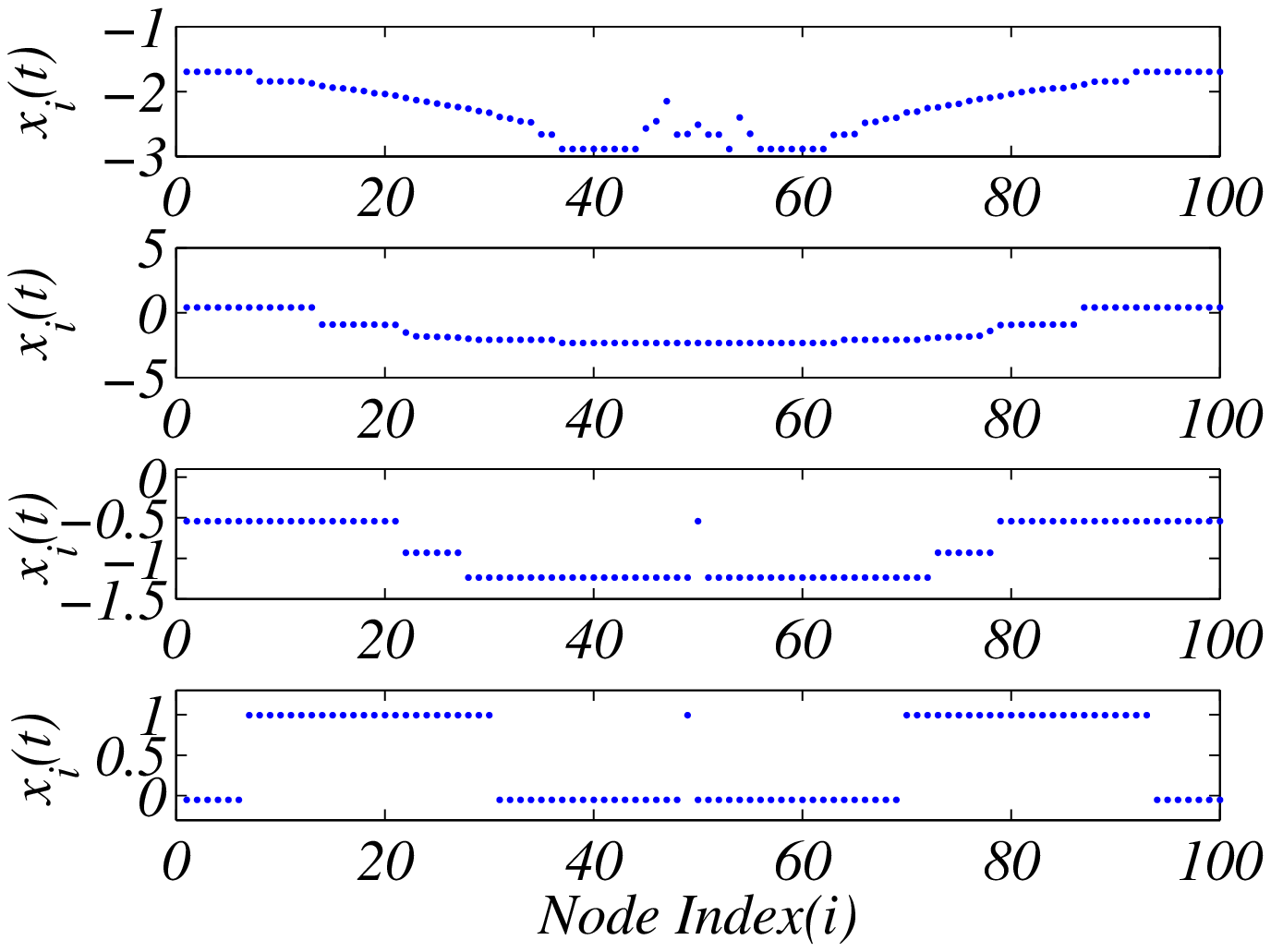}
\caption{Transition from a single cluster state to chimera states in VDP network. (a) Temporal  evolution of each node plotted for four different cross-coupling strength $\epsilon$=-11,-8,-5,-3 (from top to bottom). (b) Snapshot of $x_i$ at a particular instant of time.} 
\label{chimera_vdp}
\end{figure*}
\par We define a cross-coupling in the real plane to construct a globally coupled network of identical oscillators, which is now expressed by $\dot{X_i}=F(X_i)+ K \Gamma g$, $F: R^m\rightarrow R^m$, the second term at right side defines the coupling where $g$ is a $m\times 1$ matrix involving the dynamical variables of the network and $\Gamma$ is a $m\times m$ real matrix and $K$ is a coupling constant. 
We first consider a globally coupled network of identical VDP oscillators under such a planar type cross-coupling with the following model equations,
 \begin{subequations}
\begin{eqnarray}
 \dot{x_i}&=&y_i +K[\bar {x}-x_i -\epsilon (\bar {y}-y_i)]\\ 
 \dot{y_i}&=&b(1-x_{i}^2)y_i-x_i+K[\bar {y}-y_i +\epsilon (\bar {x}-x_i)]. \ 
\end{eqnarray}
\end{subequations}
where $\bar {x}$=$\frac{1}{N} \sum\limits_{j=1}^N {x_j}$ and $\bar {y}$= $\frac{1}{N} \sum\limits_{j=1}^N  {y_j}$ and the coupling matrix $\Gamma$ and $g$ are obtained from Eq. (1), 
\[ \Gamma=\begin{pmatrix}
  1 & -\epsilon \\
   \epsilon & 1
 \end{pmatrix};
 g=\begin{pmatrix}
   \bar{x}-x_i\\
   \bar{y}-y_i \\
 \end{pmatrix}
\]
and $\epsilon$ is the cross-coupling strength. To the conventional global mean-field coupling (involving similar variables of the system) whose strength is controlled by $K$, we have added a cross-coupling term, that employs the other variables, in a linear form and its strength is, particularly, tunable by the  parameter $\epsilon$. Such a coupling typically appears in a  hyper-network structure \cite{Sorrentino, Ram}. A similar form of cross-coupling between the activator and the inhibitor variables was also used earlier \cite{Omelchenko-2013} in a network of FitzHugh-Nagumo oscillators \cite{FitzHugh} to observe multichimera states. However, there the coupling had a nonlocality feature. 
In contrast, we maintain a global coupling without any spatial variation. In our coupling, the matrix $\Gamma$ operates on the vector $g$ when the rotational symmetry \cite{Icke}  is lost for nonzero $\epsilon$ values. The interplay of the global mean-field coupling and the linear cross-coupling is crucial for the emergence of chimera states. 
\begin{figure*} [t!]
{\includegraphics[height=7cm,width=9cm]{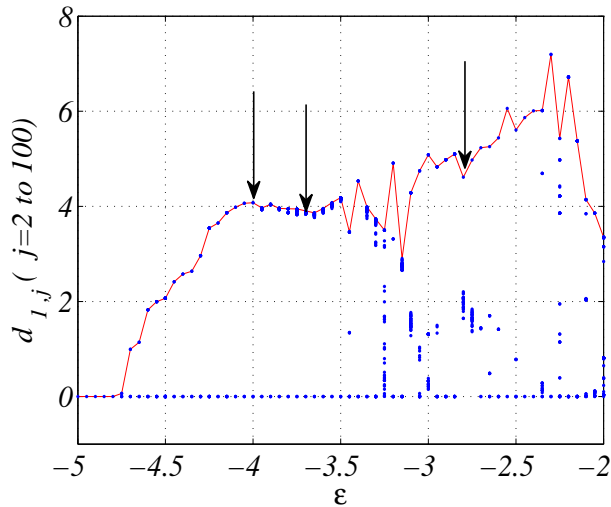}}
\caption{Network of R\"ossler oscillators. Distance $d_{1,j}$ of oscillator-1 from all others with $\epsilon$. Bifurcation from a single cluster to two cluster and chimera and multichimera states.} 
\label{distance_rossler} 
\end{figure*}
\par In a parameter space around $b=3$ ($K=0.04$), when the oscillators behave like relaxation oscillators, we observe a clear signature of chimera states in a network of size $N=100$. We use MatLab2008 and ODE45 routine for numerical check and discard 100000 data points as transients. The initial states for $y_i$ are chosen as ($y_{i0} =A\times(\frac{N}{2} -i)$ for $i=1$ to $\frac{N}{2}$) and  ($y_{i0}=A\times(i-\frac{N}{2}$) for ($i=\frac{N}{2}+1$ to N)  with an added random fluctuation and $A=0.1 $. A sequence of emergent behaviors, clustering and chimera states are observed for varying $\epsilon$. We introduce a measure consisting of the time average of the synchronization error  or the euclidean distance of any arbitrarily chosen reference oscillator (say, 1) from all the other oscillators ($j$=2, ...N) as, $d_{1,j}=<||{\vec{X_j}-\vec{X_1}}||>$, where $<>$ signifies the time average and $\vec{X_j}$ is the state vector at the $j$-th node and this is plotted in Fig. \ref{distance_vdp}. It clearly shows 1-cluster which bifurcates to 2-cluster ($\epsilon=-2.6$), 3-cluster ($\epsilon=-4.6$) and then to multi-cluster states and finally, to chimera states for varying $\epsilon$ keeping $K=0.04$ constant. The multicluster states exist for intermediate $\epsilon$ values between 3-cluster and chimera states, however, the chimera states are not clearly evident from the bifurcation diagram. The chimera states are clear from the snapshots of $x_i$ in Fig. \ref{chimera_vdp} (b). The sequence of behaviors (1-cluster, 2-cluster, multicluster and chimera) is also demonstrated in Fig. \ref{chimera_vdp}(a)  that shows the temporal evolution of $x_i$ variable of  all the nodes in each panel for various $\epsilon$ values. The lowest panel shows two coherent clusters for a cross-coupling strength $ \epsilon=-3$ followed by the upper panels with three cluster states for $\epsilon=-5$ and multi-cluster states for $\epsilon=-8$ respectively. Finally the uppermost panel (for $\epsilon=-11$) clearly shows a chimera state. 
For further evidence of the chimera states, we present snapshots of $x_{i}$ (i=100), in Fig. \ref{chimera_vdp}(b), for different coupling strengths in four panels which correspond to the immediate left panels in Fig. \ref{chimera_vdp}(a). At the bottom panel, we observe a two cluster state, and then in the two upper panels, three clusters, multicluster and at the topmost panel we find the chimera state. We note that, in a chimera state, we observe two subpopulations with coherence and incoherence in their amplitude dynamics. We have checked that this chimera feature persists for networks of larger sizes.
\par Once the chimera is confirmed in a network of the relaxation oscillator, we proceed to check if our choice of coupling can succeed in creating a chimera  in a network of chaotic units. We apply the same planar type cross-coupling to construct the  globally coupled network of R\"ossler oscillators, 
\begin{figure*} 
\subfigure(a){\includegraphics[height=9.25cm,width=7cm]{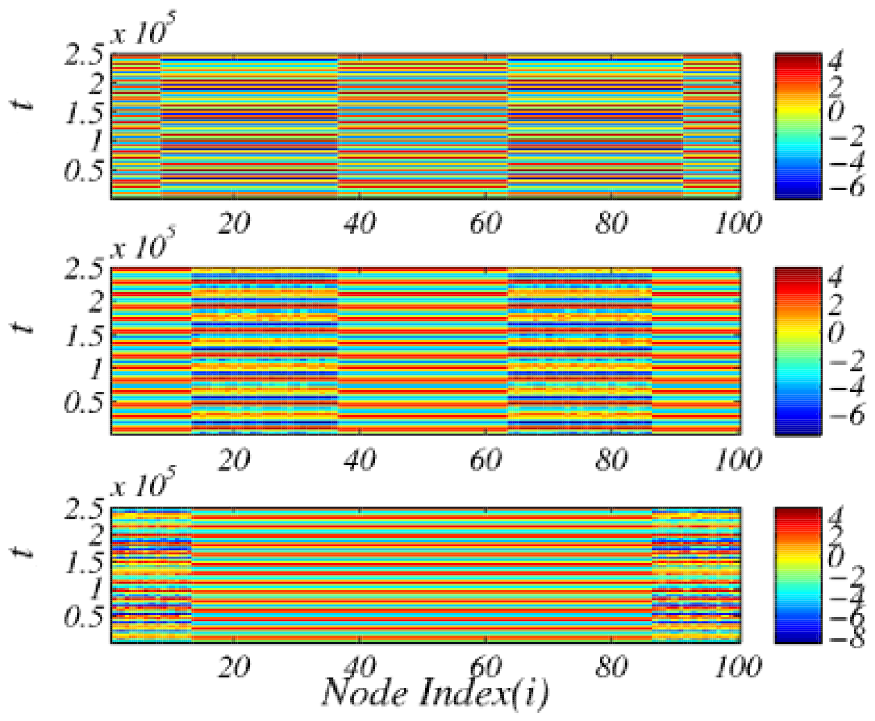}}
\subfigure(b){\includegraphics[height=9.5cm,width=7cm]{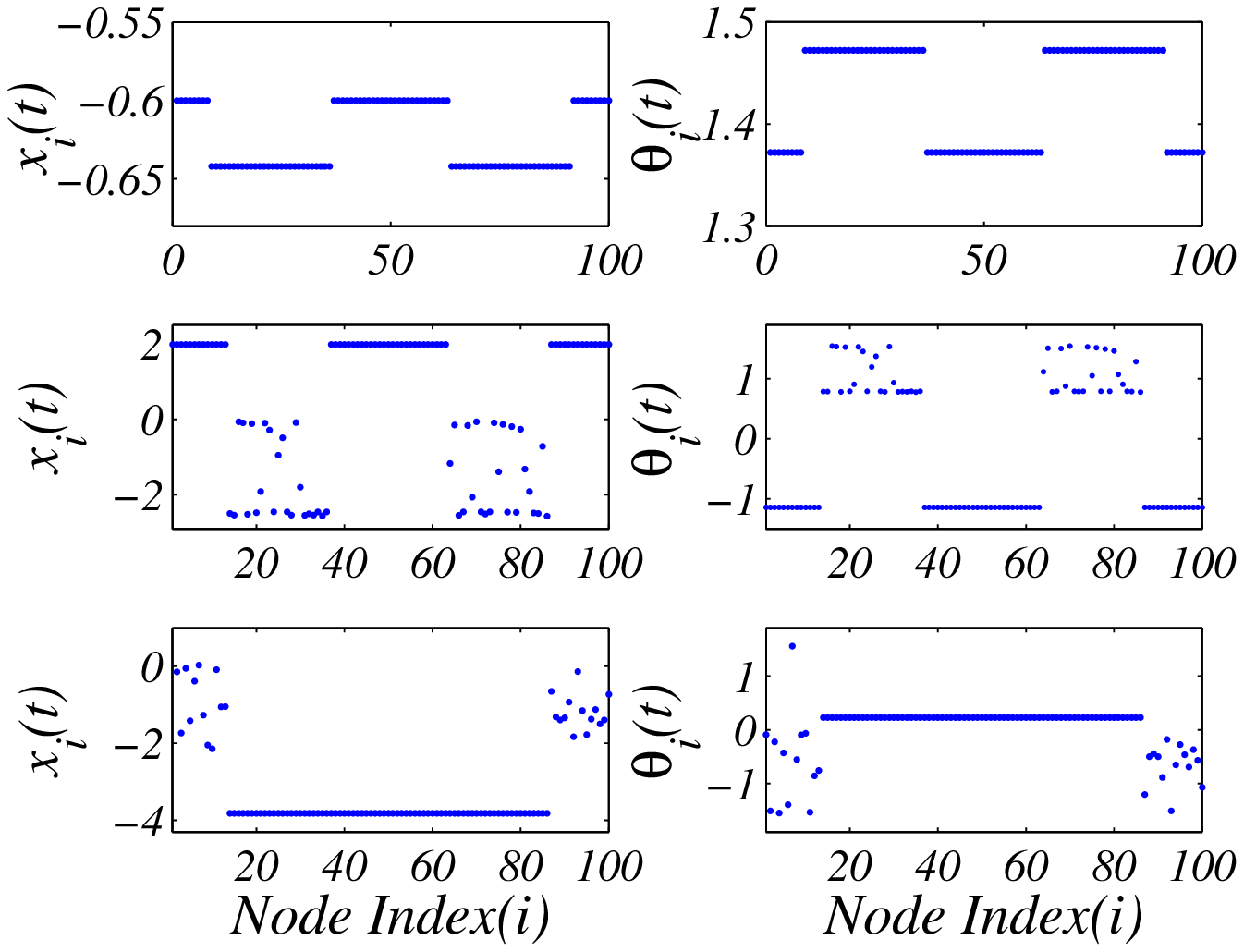}}
\caption{Multicluster and multichimera states in a globally coupled R\"ossler oscillators. 
(a) Time series $x_i$ for all nodes $i$=1 to 100, (b) snapshots of $x_i$ at left for all nodes for $K$=0.08 and $\epsilon$ =-4.0, -3.7, -2.8 respectively from
 bottom to top panels, similar snapshots of phase $\theta_i$ at right panels.}  
\label{chimera_rossler}
\end{figure*}
 \begin{eqnarray}
 \dot{x_i}&=&-y_i-z_i +K[\bar {x}-x_i -\epsilon (\bar {y}-y_i)] \ \nonumber\\
 \dot{y_i}&=&x_i+a y_i+K[\bar {y}-y_i +\epsilon (\bar {x}-x_i)]\\ \nonumber\
 \dot{z_i}&=&b x_i+z_i(x_i-c). \nonumber
\end{eqnarray}
 For this 3D system,
\[ \Gamma=\begin{pmatrix}
   0 & 0 & 0
 \end{pmatrix};
 g=\begin{pmatrix}
   \bar{x}-x_i\\
  \bar{y_j}-y_i  \\
  0
 \end{pmatrix}
\]
The $\Gamma$ matrix operates here on the  $g$ vector to break the rotational symmetry for nonzero $\epsilon$ values. The R\"ossler system parameters are taken in the chaotic regime ($a=0.36,b=0.4,c=4.5$). We apply three different cross-coupling strengths ($\epsilon= -4.0,-3.7,-2.8$) and $K=0.08$ when we identify two  cluster, multicluster chimera and single chimera regimes. This sequence of emergent behaviors, particularly, the multi-chimera state under global coupling is a new feature. To characterize the clustering and chimera from all the other oscillators ($j$=2, ...N), we use the same euclidean distance measure $d_{1,j}$ as described above.  The $d_{1,j}$ of each node of the network (N=100) is plotted in Fig. \ref{distance_rossler} as a function of the cross-coupling strength $\epsilon$. For $\epsilon$= -4.75 and below, the $d_{1j}$ plot of all the nodes confirms a single cluster: all oscillators are completely synchronized. With a slight increase in $\epsilon$ the whole population splits into two synchronized clusters. In contrast to the common notion of attaining a symmetry, an inhomogeneity is created with increasing $\epsilon$. A similar inhomogeneity with increasing coupling strength was reported \cite{Daido} in a globally coupled CGLE system, however, they did not notice any chimera state. The coupling plays an important role in our case where it is not a simple all-to-all global coupling but a mixed cross-coupling. As a result, we record a splitting of a single cluster to two cluster states with increasing inhomogeneity and finally observe chimera and multichimera states for larger coupling strengths. The first arrow (Fig. \ref{distance_rossler}) from left at $\epsilon=-4.0$ confirms one of such two cluster states. Next, at $\epsilon=-3.7$, as indicated by the middle arrow, $d_{1,j}$ is measured where one node shows zero value while other nodes are scattered with finite values, however, this is not so clear in this Fig. \ref{distance_rossler}, which is actually a multichimera state but clarified below by a snapshot plot of $x_i$ in Fig.~\ref{chimera_rossler}. Another example (right arrow) has been taken at $\epsilon=-2.8$ where it confirms the existence of chimera state. Numerical distance ($d_{1,j}$) is determined taking  250000 data points after discarding 200000 data points.  
\par To further clarify the collective behavior in the chaotic R\"ossler network, we plot the $x_i$ variable of all the nodes in Fig. \ref{chimera_rossler}(a).
 Color coding of the top left panel shows two clustered states for $\epsilon=-4.0$, when the middle panel exhibits multicluster chimera states ($\epsilon=-3.7$) and the lower panel shows single chimera (middle region is in coherent). Furthermore, we take snapshots of $x_i$ and phase $\theta_i$ at an instant of time  in Fig. \ref{chimera_rossler}(b) for all the 100 oscillators. They reconfirm the existence of two cluster, multichimera and single chimera states in top to bottom panels corresponding to $\epsilon=-4.0, -3.7, -2.8$ respectively. A similar two cluster and multiclustered and single clustered chimera is also reflected in the snapshots of instant phase $\phi_i$ for all the oscillators at right panels. The last two panels (middle and lower) at right in Fig. \ref{chimera_rossler}(b) exhibit coexistence of randomly distributed phases with coherent phases leading to multi-chimera and chimera states respectively.
 To summarize, we have observed chimera states in networks of VDP oscillators as well as chaotic R\"ossler oscillators using a planar type global coupling. For the case of the van der Pol oscillators the evolutionary path to the chimera state follows a sequence of a single cluster state to a two cluster state and then a chimera state as the cross-coupling strength $\epsilon$ is varied. For the globally coupled R\"ossler system in the chaotic regime we observe both chimera and multi-chimera states. The chimera states for both the oscillator systems show amplitude and phase fluctuations in the incoherent part of the subpopulation while they are coherent in the other subpopulation of the network. 
It is evident that nonisochronicity \cite{Mauroy, Kopell, Daido, Blasius} is an important ingredient for the emergence of chimera in limit cycle systems such as VDP system under global coupling while the rotational symmetry breaking \cite{Icke} plays additional role. Non-isochronicity in limit cycle systems allows amplitude fluctuations in the non-coherent population in the chimera state. For the R\"ossler system operating in the chaotic regime, amplitude fluctuations are intrinsic to the dynamics where the presence of non-isochronicity is redundant. 
It would be worth testing out this coupling in a host of other systems to verify the conceptual basis of this mechanism. Since the  linear cross-coupling form is easy to implement it can also be tried out experimentally and provide a useful paradigm for a broad range of applications of the chimera state.

\par C.R.H., S.K.D. and P.K.R. acknowledge support by the CSIR Emeritus scientist schemes. A. Mishra is supported by the UGC-NET Fellowship.

\end{document}